\documentclass[aps,prl,floatfix,twocolumn,superscriptaddress]{revtex4-2}
\usepackage[utf8]{inputenc}
\usepackage{amsmath,amssymb,amsfonts,amssymb,bm}
\usepackage{physics}
\usepackage{graphicx,color,soul}
\graphicspath{{plots/}}
\usepackage[caption=false]{subfig}
\usepackage{tabularx,booktabs}
\usepackage{tikz}
\usetikzlibrary{arrows.meta}
\usepackage{hyperref}
\usepackage{enumerate}
\makeatletter
\def\input@path{{figs/}}
\makeatother

\begin{document}

\title{Magnetopological mechanics in Maxwell lattice frustrated Mott insulators}
\author{Hong-Hao Song}
\affiliation{International Center for Quantum Materials, School of Physics, Peking University, Beijing 100871, China}
\author{Pengwei Zhao}
\affiliation{International Center for Quantum Materials, School of Physics, Peking University, Beijing 100871, China}
\author{Gang v.~Chen}
\email{chenxray@pku.edu.cn}
\affiliation{International Center for Quantum Materials, School of Physics, Peking University, Beijing 100871, China}
\affiliation{Collaborative Innovation Center of Quantum Matter, 100871, Beijing, China}
\date{\today}

\begin{abstract}
Topological boundary modes, a hallmark of quantum topological phases, remarkably 
occur in classical mechanical systems through an interesting correspondence 
with the quantum case. Here, we explore the Maxwell lattice frustrated Mott insulators 
and argue that the combination of the intrinsic spin-lattice coupling and the spin exchanges 
could induce the topological mechanics with topological boundary floppy modes 
in the phonon spectra. This mechanism and phenomena are dubbed magnetic topological mechanics, or, 
magnetopological mechanics in short. Focusing on a two-dimensional kagom\'e lattice spin model, 
we illustrate how strong spin-lattice coupling drives a spontaneous lattice distortion, 
resulting in the topological Maxwell lattice with the topological polarization and non-trivial phonon spectra.  
Moreover, the magnetic field, that directly changes the spin state, indirectly influences
the lattice structure via the spin-lattice coupling, thereby providing a method to control the 
Maxwell lattice and the boundary modes.   
We expect this work to inspire interests in the Maxwell lattice Mott insulating materials 
and the coupling between lattices and electronic orders. 
\end{abstract}

\maketitle

\noindent{\it Introduction.}---Topological phases of matter often host exotic gapless boundary 
excitations that are robust against local perturbations, reflecting the bulk-boundary correspondence. 
Prominent examples include the chiral edge modes in quantum Hall 
effects~\cite{PhysRevB.25.2185,PhysRevB.43.11025,PhysRevB.41.12838} 
and the helical edge modes in quantum spin Hall insulators~\cite{PhysRevLett.95.226801,PhysRevLett.96.106802}. 
Interestingly, analogous phenomena can also arise in classical mechanical systems, 
where the dynamical matrix intriguingly plays the role of an effective Hamiltonian 
and enables a topological classification of phonon bands. 
Within this framework, the Maxwell mechanical structures with the non-trivial 
topological polarization can host topological boundary modes enforced by the bulk topology, 
which are likewise robust against local perturbations~\cite{KaneLubensky2014,Huber2016}.

A Maxwell lattice is defined as a system where the number of constraints (springs) balances 
the number of degrees of freedom (coordinates of sites). 
Kagom\'e lattice in two dimensions and pyrochlore lattice in three dimensions 
are well-known Maxwell lattices.
Cutting a finite portion of such lattices removes constraints of the systems 
and necessarily introduces zero modes localized at the boundaries. 
Within the phononic topological band theory framework, these zero modes are 
classified into local floppy modes and topological floppy modes.
The former is determined from the traditional Maxwell Index theorem~\cite{Maxwell1864,CALLADINE1978161,Nakahara2018}, 
while the latter depends on the bulk topological polarization~\cite{KaneLubensky2014} 
and can be used to label distinct mechanical phases.

While these intriguing boundary topological floppy modes have been demonstrated 
in artificial spring and mass mechanical networks, we raise a natural question 
that is weather such topological mechanical phenomena emerge naturally in 
quantum materials without deliberately engineered architectures. 
The observation is from the fact that many representative Maxwell lattices 
like kagom\'{e} and pyrochlore lattices are frustrated lattices 
where many frustrated spin physics were studied. 
To generate the topological boundary modes, 
one has to convert the perfect lattice 
into the topological Maxwell lattice in these materials. 
A promising route to achieving this lies in the intrinsic 
spin-lattice coupling (SLC)~\cite{PhysRevLett.93.197203,PhysRevB.74.134409,PhysRevLett.88.067203,PhysRevB.66.064403,PhysRevB.105.174424}.
Through SLC, these systems may undergo spontaneous lattice distortions 
that can potentially reduce or even remove the non-trivial states of self stress (SSSs), 
thereby possibly driving the structure toward a topological Maxwell lattice 
with a non-trivial bulk topological polarization. 
Its phonon spectrum is expected to exhibit the corresponding 
topologically-protected floppy modes.

\begin{figure}[b]
    \centering
    \subfloat{%
    \begin{tikzpicture}
    \node[anchor=north west, inner sep=0] (img) at (0,0)
        {\includegraphics[width=0.47\columnwidth]{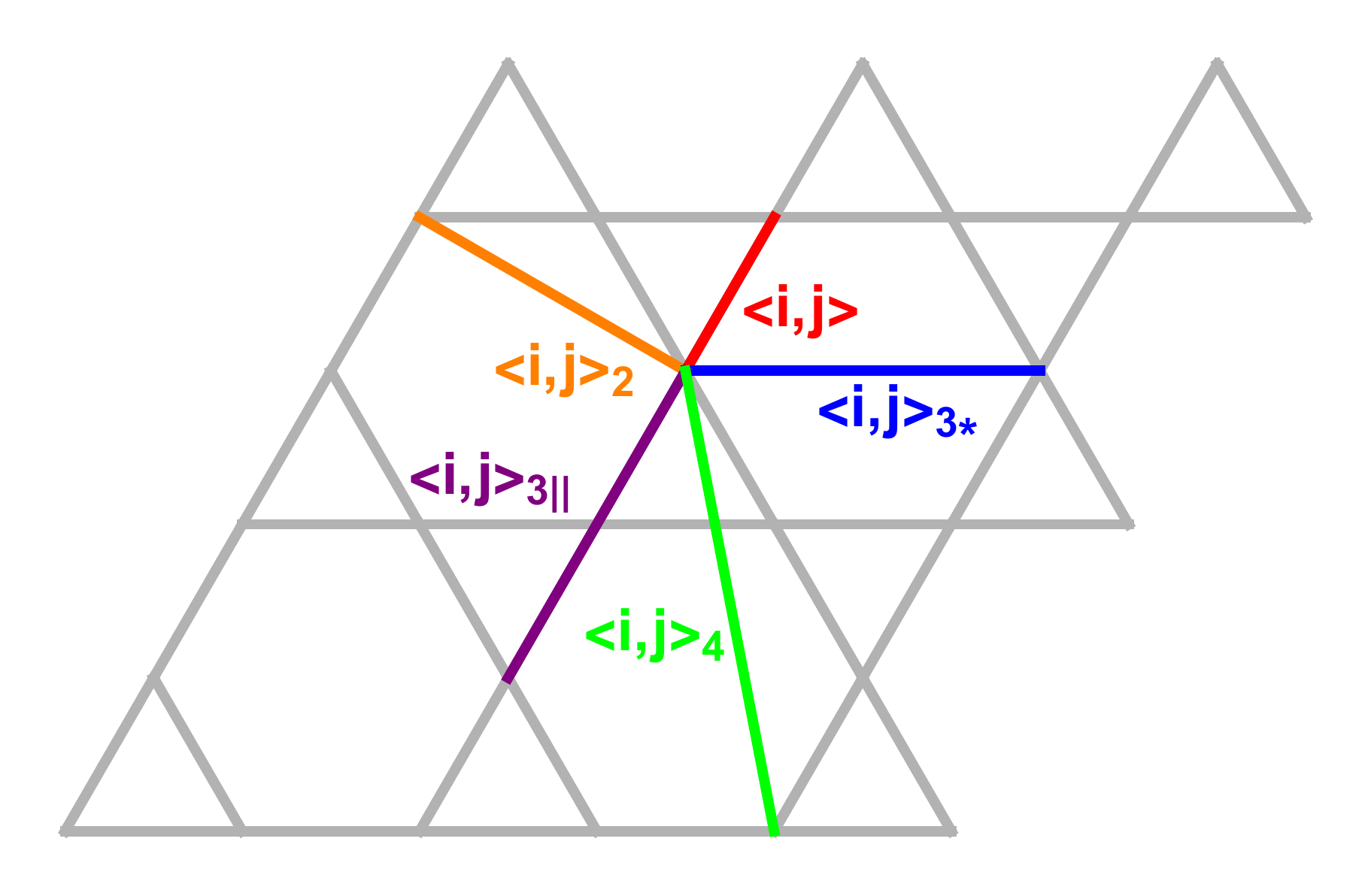}};
    \node[anchor=north west] at ([xshift=1pt,yshift=8pt]img.north west)
        {\fontsize{6}{12}\selectfont\textbf{(a)}};
    \end{tikzpicture}%
    \label{fig:interaction}%
    }%
    \hfill
    \subfloat{%
    \begin{tikzpicture}
    \node[anchor=north west, inner sep=0] (img) at (0,0)
        {\includegraphics[width=0.5\columnwidth]{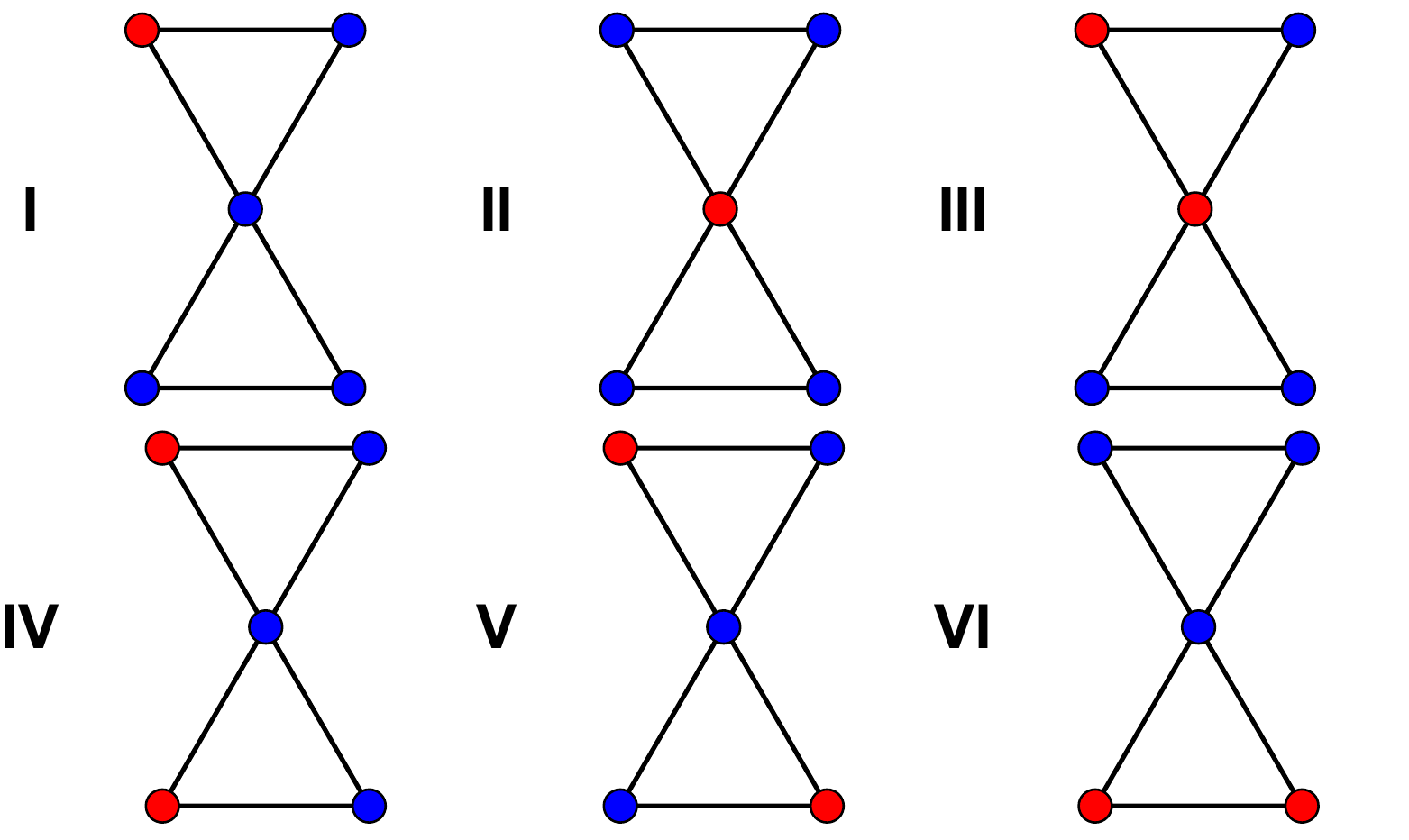}};
    \node[anchor=north west] at ([xshift=1pt,yshift=8pt]img.north west)
        {\fontsize{6}{12}\selectfont\textbf{(b)}};
    \end{tikzpicture}%
    \label{fig:local_spin_configuration}%
    }%

    \caption{(a) The spin interactions on the kagom\'e lattice.
        (b) Representative local spin configurations. 
        Red and blue circles refer to anti-aligned spins with spin up and down, respectively. 
        These configurations, together with their symmetry-related counterparts,
        act as ``Lego blocks'', from which arbitrary global spin configurations can be constructed.}
    \label{fig:side_by_side}
\end{figure}

In this Letter, we introduce the notion of ``magnetopological mechanics''
through a concrete example on a kagom\'e lattice Mott insulator with 
a strong SLC, where
we show that the SLC can induce topological mechanics. 
Specifically, we investigate a classical Heisenberg model with 
competing exchange interactions. 
It is shown that, the strong SLC regime breaks the spin degeneracy
and drives the system into a magnetically ordered state. The resulting distorted 
lattice structure eliminates the non-trivial SSSs and 
possesses a phonon spectrum characterized by the topological polarization. 
The external magnetic field is further shown to control the Maxwell lattice
 indirectly through the Zeeman coupling and the SLC. 
 The magnetopological mechanics can be generally found in many other 
 Maxwell Mott insulating materials.

\noindent{\it Spin-lattice coupling.}---To establish how topological floppy modes 
emerge microscopically in magnetic Mott insulators, 
we first introduce the SLC and demonstrate 
that it drives the lattice distortions, with the potential to produce a topological lattice.  
We consider the classical Heisenberg model and 
incorporate the magnetoelastic coupling, 
\begin{equation}
    \mathcal{H}
    = \sum_{i,j} J_{ij} \, \mathbf{S}_i \cdot \mathbf{S}_j
    + \frac{k}{2} \sum_i |\mathbf{u}_i|^2,
    \label{eq:H_KHAF}
\end{equation}
where the sum over $i,j$ runs over all interacting pairs.
Here, $\mathbf{S}_i$ represents a classical spin of unit magnitude 
($|\mathbf{S}_i|=1$), $\mathbf{u}_i$ denotes the displacement of site $i$ 
from its equilibrium position $\mathbf{R}_i$, 
with ${k>0}$ the corresponding elastic spring constant. 
In Eq.~\eqref{eq:H_KHAF}, we have adopted the site phonon model instead of the bond phonon model~\cite{PhysRevLett.93.197203,PhysRevB.74.134409,PhysRevB.66.064403,PhysRevLett.88.067203},
and the latter is actually simpler in the effective spin model but does not directly provide 
 simple expressions for the site displacement.  
Although these two modes are quite different, the number of degrees of freedom is 
identical on the Maxwell lattice. 
$J_{ij}$ is the exchange interaction, 
which is assumed to isotropically depend on the bond length as 
$J_{ij} = J(|\mathbf{R}_{ij} + \mathbf{u}_i - \mathbf{u}_j|)$,
where $\mathbf{R}_{ij} = \mathbf{R}_i - \mathbf{R}_j$.
For small displacements $|\mathbf{u}_i|/|\mathbf{R}_i| \ll 1$, 
the exchange interaction can be expanded linearly as
\begin{equation}
    J_{ij} \approx J + \left( \frac{\mathrm{d}J_{ij}}{\mathrm{d}r} \right)_{R = |\mathbf{R}_{ij}|} 
    \hat{\mathbf{e}}_{ij} \cdot (\mathbf{u}_i - \mathbf{u}_j),
    \label{eq:J_expand}
\end{equation}
where ${\hat{\mathbf{e}}_{ij} = \mathbf{R}_{ij} / |\mathbf{R}_{ij}|}$ 
is the unit vector along the bond.
This so-called exchange striction couples the local site distortion 
$\mathbf{u}_i$ to the spin interaction 
${\mathbf{S}_i \cdot \mathbf{S}_j}$. 
For the $n$th-neighbor exchange interaction, the SLC enters through 
the derivative $\nabla J_n \equiv (\mathrm{d}J_n/\mathrm{d}R)_{R=|\mathbf{R}_{ij}|}$.
The relative SLC strength of the $n$th-neighbor term compared to 
the nearest-neighbor (NN) one is therefore characterized by $|\nabla J_n/\nabla J_1|$.
In the realistic settings, exchange couplings decay rapidly with distance, 
so that ${|J_n| \ll |J_1|}$ for ${n\ge 2}$, and correspondingly 
${|\nabla J_n/\nabla J_1| \ll 1}$. We thus only retain the dominant 
SLC contribution arising 
from the NN interaction in the following analysis.

An effective spin Hamiltonian is obtained by integrating out the lattice degrees of freedom.
Since the Hamiltonian is quadratic in the site displacements, 
this Gaussian integration is equivalent to minimizing $\mathcal{H}$ 
with respect to the lattice degrees of freedom ${\mathbf{u}_i}$.
This procedure yields the optimal site displacements
\begin{equation}
    \mathbf{u}_i^{*}
    = -\frac{J \gamma }{k}\sum_{j \in \mathcal{N}(i)}
    (\mathbf{S}_i \cdot \mathbf{S}_j)\, \hat{\mathbf{e}}_{ji},
    \label{eq:u_opt}
\end{equation}
where ${J= J_1(|\mathbf{R}_{ij}|)}$, 
${\gamma=\frac{1}{J}(\mathrm{d}J_1/\mathrm{d}R)|_{R=|\mathbf{R}_{ij}|}}$ 
and $\mathcal{N}(i)$ denotes the set of four NN sites of $i$.
Eq.~\eqref{eq:u_opt} shows explicitly that the lattice distortion 
at site $i$ is driven by the imbalance of spin bond energy around the site.
Therefore, determining the distorted lattice requires prior knowledge 
of the underlying spin configuration. 
Since the latter is selected self-consistently by the SLC, 
we first integrate out the lattice degrees of freedom to obtain an effective spin Hamiltonian.
Substituting Eq.~\eqref{eq:u_opt} back into Eq.~\eqref{eq:H_KHAF}, we obtain 
the effective spin model,
\begin{equation}
    \mathcal{H}_{\mathrm{eff}}
    = \sum_{\langle i,j \rangle_n} J_{n}
      \mathbf{S}_i \cdot \mathbf{S}_j 
      - J b \sum_{\langle i,j \rangle_1} (\mathbf{S}_i \cdot \mathbf{S}_j)^2 
      + \mathcal{H}_{\mathrm{FN}},
    \label{eq:H_eff}
\end{equation}
where $J_n$ represents the $n$th-neighbor interaction ($J_1=J$) and 
$b = \frac{J \gamma^2}{k}$ is a dimensionless SLC strength.
The three-spin quartic term $\mathcal{H}_{\mathrm{FN}}$ takes the form
\begin{equation}
    \mathcal{H}_{\mathrm{FN}}
    = - \frac{Jb}{2} 
      \sum_i \sum_{j \neq k \in \mathcal{N}(i)} 
      (\hat{\mathbf{e}}_{ij} \cdot \hat{\mathbf{e}}_{ik})
      (\mathbf{S}_i \cdot \mathbf{S}_j)
      (\mathbf{S}_i \cdot \mathbf{S}_k).
    \label{eq:H_FN}
\end{equation}
By minimizing the effective Hamiltonian $\mathcal{H}_{\mathrm{eff}}$, 
one obtains the ground spin configurations, 
which in turn determine the lattice distortion via Eq.~\eqref{eq:u_opt}.

\noindent{\it SLC-induced topological lattice.}---To realize a topological Maxwell lattice 
whose phonon spectrum carries a nontrivial topological polarization, 
we consider an extended Heisenberg model,
\begin{equation}
\begin{aligned}
    \mathcal{H}_{ex}
    = & \sum_{\langle i,j \rangle} J_{ij} \, \mathbf{S}_i \cdot \mathbf{S}_j
    + \sum_{\langle i,j \rangle_{2}} J_{2} \, \mathbf{S}_i \cdot \mathbf{S}_j 
   + \sum_{\langle i,j \rangle_{3\parallel}} J_{3\parallel} \, \mathbf{S}_i \cdot \mathbf{S}_j\\
     & + \sum_{\langle i,j \rangle_{3*}} J_{3*} \, \mathbf{S}_i \cdot \mathbf{S}_j
    + \sum_{\langle i,j \rangle_{4}} J_{4} \, \mathbf{S}_i \cdot \mathbf{S}_j,
    \label{eq:H_extended}
\end{aligned}
\end{equation}
where ${J_{ij}>0}$, $J_{2}$, $J_{3\parallel}$, $J_{3*}$ and $J_4$ denote 
the NN, 2nd neighbor, bond-directional 3rd neighbor, 
plaquette-crossing 3rd neighbor 
and 4th neighbor exchange interactions, respectively, 
as illustrated in Fig.~\ref{fig:interaction}.

We then determine the SLC-induced lattice distortion 
by analyzing the ground state configurations of the effective Hamiltonian~\eqref{eq:H_eff}.
In the NN limit (${J_n =0 }$ for ${n>1}$), 
the strong SLC (${b>1/6}$) favors 
the collinear spin configurations 
due to the biquadratic spin interaction 
in Eq.~\eqref{eq:H_eff}~\cite{PhysRevB.105.174424}. 
Within the collinear spin configurations, 
the effective Hamiltonian can be mapped onto an Ising-type model 
by setting ${\mathbf{S}_i = \hat{z}\,\sigma_i}$ with ${\sigma_i = \pm 1}$. 
Since the further-neighbor exchanges are much weaker 
than the NN coupling in realistic settings, the Ising description 
remains valid in the strong SLC regime, 
leading to an Ising-type effective Hamiltonian,
\begin{equation}
\begin{aligned}
    \mathcal{H}_{ex}^{\mathrm{Ising}}
    = & \sum_{\langle i,j \rangle} \big(J - \frac{Jb}{2}\big) \sigma_i \sigma_j
    + \sum_{\langle i,j \rangle_{2}} \big(J_2 + \frac{Jb}{2}\big) \sigma_i \sigma_j \\
   &+ \sum_{\langle i,j \rangle_{3\parallel}} \big(J_{3\parallel} + Jb \big) \sigma_i \sigma_j
    + \sum_{\langle i,j \rangle_{3*}} J_{3*}\, \sigma_i \sigma_j\\
    &+ \sum_{\langle i,j \rangle_{4}} J_{4}\, \sigma_i \sigma_j .
    \label{eq:H_extended_Ising}
\end{aligned}
\end{equation}

When ${J_{3*}=0}$ and ${J_{4}=0}$, the ground state configurations 
can be determined by several local spin patterns on the ``bow-tie'' unit, 
as illustrated in Fig.~\ref{fig:local_spin_configuration}.
The energies of these local spin patterns are
$E^{(\mathrm{I})}=\frac{2}{3}J$,
$E^{(\mathrm{II})}=2J_2+2J_{3\parallel}+2Jb-\frac{2}{3}J$,
$E^{(\mathrm{III})}=-\frac{2}{3}J$, 
$E^{(\mathrm{IV})}=2J_2-2J_{3\parallel}-\frac{2}{3}J$,
$E^{(\mathrm{V})}=-2J_2+2J_{3\parallel}+2Jb-\frac{2}{3}J$, and
$E^{(\mathrm{VI})}=-2J_2-2J_{3\parallel}-4Jb+\frac{2}{3}J$, and 
the details are found in the Appendix. 
All these patterns, together with their symmetry-related counterparts, 
act as ``Lego blocks'',
from which arbitrary global configurations 
on the kagom\'e lattice can be constructed.

In the NN limit, for ${1/6 < b < 1/3}$ region, 
the lowest-energy patterns are type $\mathrm{III}$, $\mathrm{IV}$, 
and their symmetry-related counterparts. 
The global ground state configurations can then be constructed 
by randomly assembling type $\mathrm{III}$ and type $\mathrm{IV}$ 
patterns with ratio ${2:1}$, leading to a macroscopically 
degenerate ground state.
Turning on the weak antiferromagnetic further-neighbor interactions, 
types~$\mathrm{III}$ and~$\mathrm{IV}$ are still the two lowest-energy local patterns, 
while which one is favored depends on the specific values of $J_2$ and $J_{3\parallel}$.
However, neither type $\mathrm{III}$ nor type $\mathrm{IV}$
can tile the kagom\'e lattice by itself; thus, the ground state configuration consists 
of mixed type $\mathrm{III}$ 
and type $\mathrm{IV}$ patterns.
According to Kanamori's method of inequalities~\cite{10.1143/PTP.35.16,MWolf_1988}, 
this ground state is in the same phase as the NN limit.
This phase is bounded by the inequalities
${J_2-J_{3\parallel}<\frac{Jb}{2}}$,
${J_2>-\frac{Jb}{2}}$, and
${J_2+2J_{3\parallel}<J-3Jb}$. 

\begin{figure}[t]
    \centering
    \includegraphics[width=0.46\textwidth]{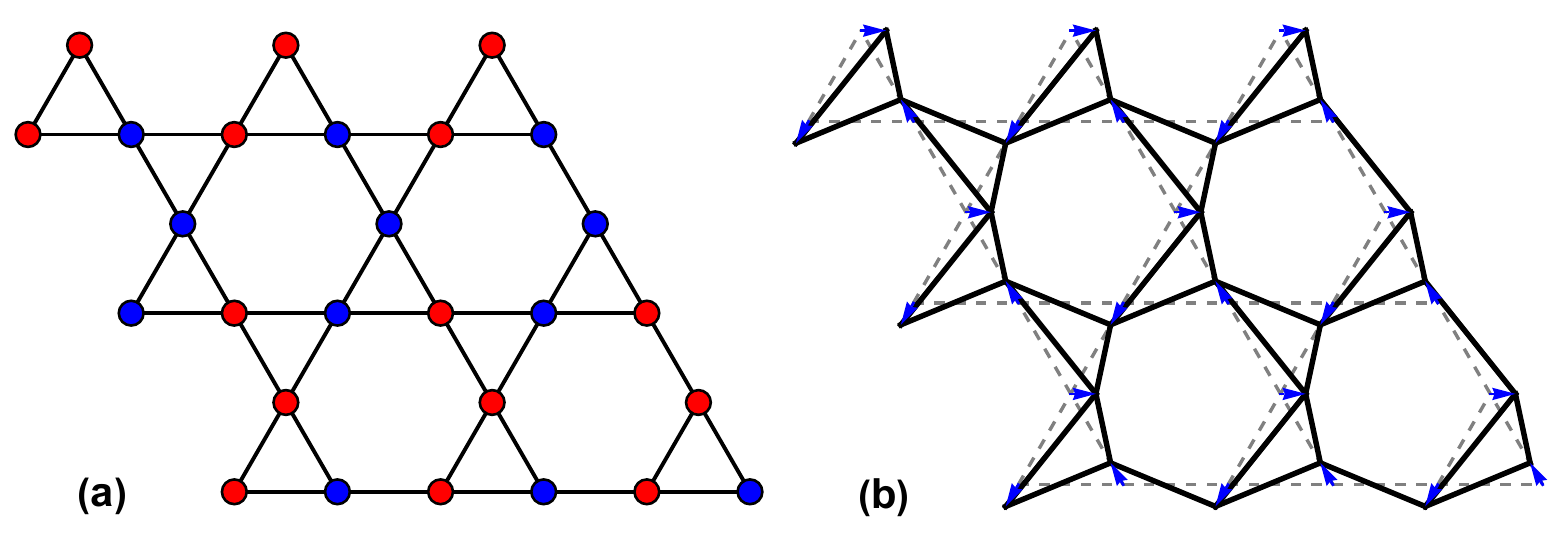}
    \caption{(a) Spin configuration obtained from simulated annealing 
    at ${J=1}$, ${J_2=0.05}$, ${J_{3||}=0.03}$, ${J_{3*}=0.03}$, ${J_4=-0.02}$, ${b=0.2}$. 
    Red circles indicate spin up ($+\hat{z}$), while blue circles indicate spin down (${-\hat{z}}$).
    (b) Corresponding lattice distortion induced by the SLC.}
    \label{fig:spin_lattice}
\end{figure}

Introducing a weak antiferromagnetic $J_{3*}$ together with a weak 
ferromagnetic $J_4$ lifts the macroscopic degeneracy and selects an ordered state.
We identify this selected order by performing simulated annealing 
with periodic boundary conditions on a ${3\times10\times10}$ lattice; 
a representative ground state spin configuration is shown in Fig.~\ref{fig:spin_lattice}(a).
The resulting phase is twelvefold degenerate, reflecting the spontaneous breaking of a 
$\mathbb{Z}_3 \times \mathbb{Z}_2 \times \mathbb{Z}_2$ 
symmetry associated with lattice rotations, inversion, and global spin inversion, respectively.
Once the ground state spin configuration is fixed, the 
SLC-induced lattice distortion follows 
directly from Eq.~\eqref{eq:u_opt} and is shown in Fig.~\ref{fig:spin_lattice}(b).
This distortion bends the originally straight kagom\'e filaments into the zigzag-like chains.
Since the SSSs in the kagom\'e lattice rely on these straight bond lines, 
such bending is expected to strongly suppress, and possibly eliminate, 
the SSSs and the corresponding bulk zero modes.

\begin{figure}[t]
    \centering
    \includegraphics[width=0.46\textwidth]{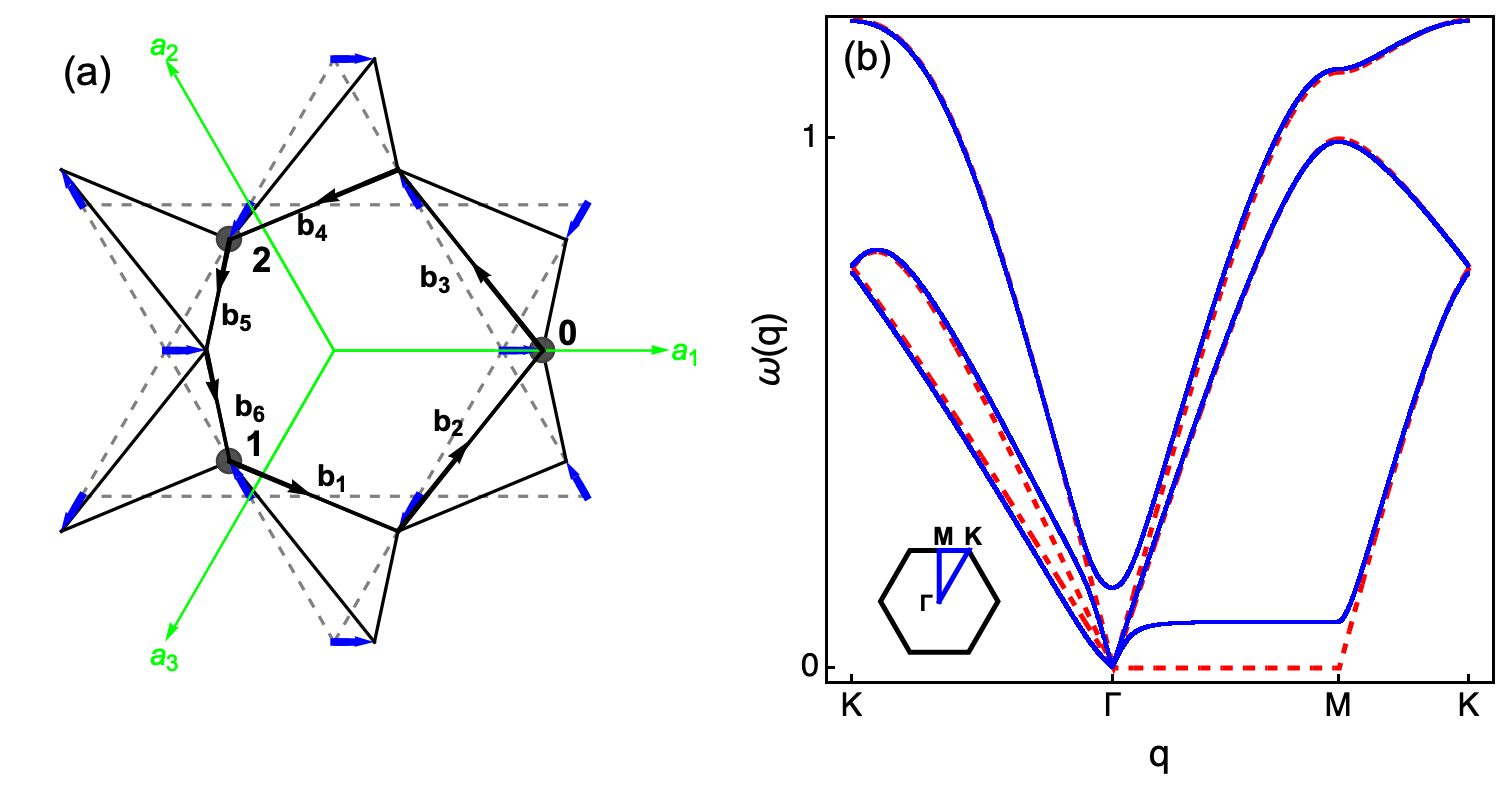}
    \caption{(a) Unit cell of the distorted kagom\'e lattice. 
    It contains three sites labelled by $0$-$2$ 
    and six bonds labelled by $b_1$-$b_6$.
    The blue arrows represent displacements of sites, 
    and green arrows $a_1$-$a_3$ are three basis vectors 
    along translational symmetric directions.
             (b) The three lowest-frequency phonon modes along $K$-$\Gamma$-$M$-$K$, 
             with the inset the Brillouin zone. 
             The red (blue) dashed lines are phonon modes of the perfect (distorted) kagom\'e lattice,  
             for $\frac{J\gamma}{k}=-0.0125$}.
    \label{fig:unit_cell}
\end{figure}

To verify whether the SLC-induced distortion indeed removes the SSSs 
and drives the lattice into a topological Maxwell lattice, 
we proceed to consider the bulk phonon spectrum under periodic boundary conditions (PBCs).
Although the spin sector helps distort the lattice and drives the formation of 
topological lattice, the lattice dynamics and 
the spin dynamics are effectively independent at the quadratic level. 
Thus, at the lowest order of approximation, 
we can start from the distorted lattice structure and directly compute 
the phonon spectrum. 
As shown in Fig.~\ref{fig:unit_cell}(a), the kagom\'e unit cell contains 
three sites labelled $0$--$2$ and six bonds labelled $\mathbf{b}_1$--$\mathbf{b}_6$.
In the absence of SLC, the bond vectors are conveniently expressed in terms of 
the basis vectors 
${\mathbf{a}_n=\bigl(\cos\frac{2\pi(n-1)}{3},\,\sin\frac{2\pi(n-1)}{3}\bigr) }$ as
${ \mathbf{b}^K=\frac{1}{2}\{\mathbf{a}_1,-\mathbf{a}_3,\mathbf{a}_2,-\mathbf{a}_1,\mathbf{a}_3,-\mathbf{a}_2\} }$.
When the SLC is present, the distortion modifies each bond according to
$\mathbf{b}_i=\mathbf{b}^K_i+\mathbf{d}_{(i+1)\bmod 3}-\mathbf{d}_{i\bmod 3}$,
where $\mathbf{d}_n=-\frac{2J\gamma}{k}\bigl(\cos\frac{2\pi}{3}n,\,\sin\frac{2\pi}{3}n\bigr)$ 
denotes the SLC-induced displacement of the sublattice site $n$ within the unit cell.
With these distorted bond vectors, 
the bulk phonon spectrum under PBCs 
can be obtained straightforwardly~\cite{doi:10.1073/pnas.1119941109}.

The three lowest-frequency phonon modes along the $K$-$\Gamma$-$M$-$K$ line are 
shown in Fig.~\ref{fig:unit_cell}(b). In the absence of SLC (red dashed lines),  
the phonon spectrum exhibits three zero modes at ${q = 0}$, 
two rigid translations and one floppy mode that alters the unit cell area only 
at second order in displacements. In addition, the kagom\'e lattice hosts zero modes 
along the three symmetry-related $\Gamma$-$M$ directions, giving a total of $3N$ zero modes 
for a ${3\times N\times N}$ system. These modes are tied to the $3N$ straight filaments, 
consistent with the Maxwell index theorem.
When SLC is switched on (blue lines), the distortion breaks the straight filaments 
and gaps out the bulk zero modes along the $\Gamma$-$M$ directions.
Thus, only the two trivial zero modes associated with rigid translations at ${q=0}$ remain. 
Therefore, SLC effectively drives this kagom\'e lattice into a topological Maxwell lattice.

\begin{figure}[t]
    \centering
    \includegraphics[width=0.46\textwidth]{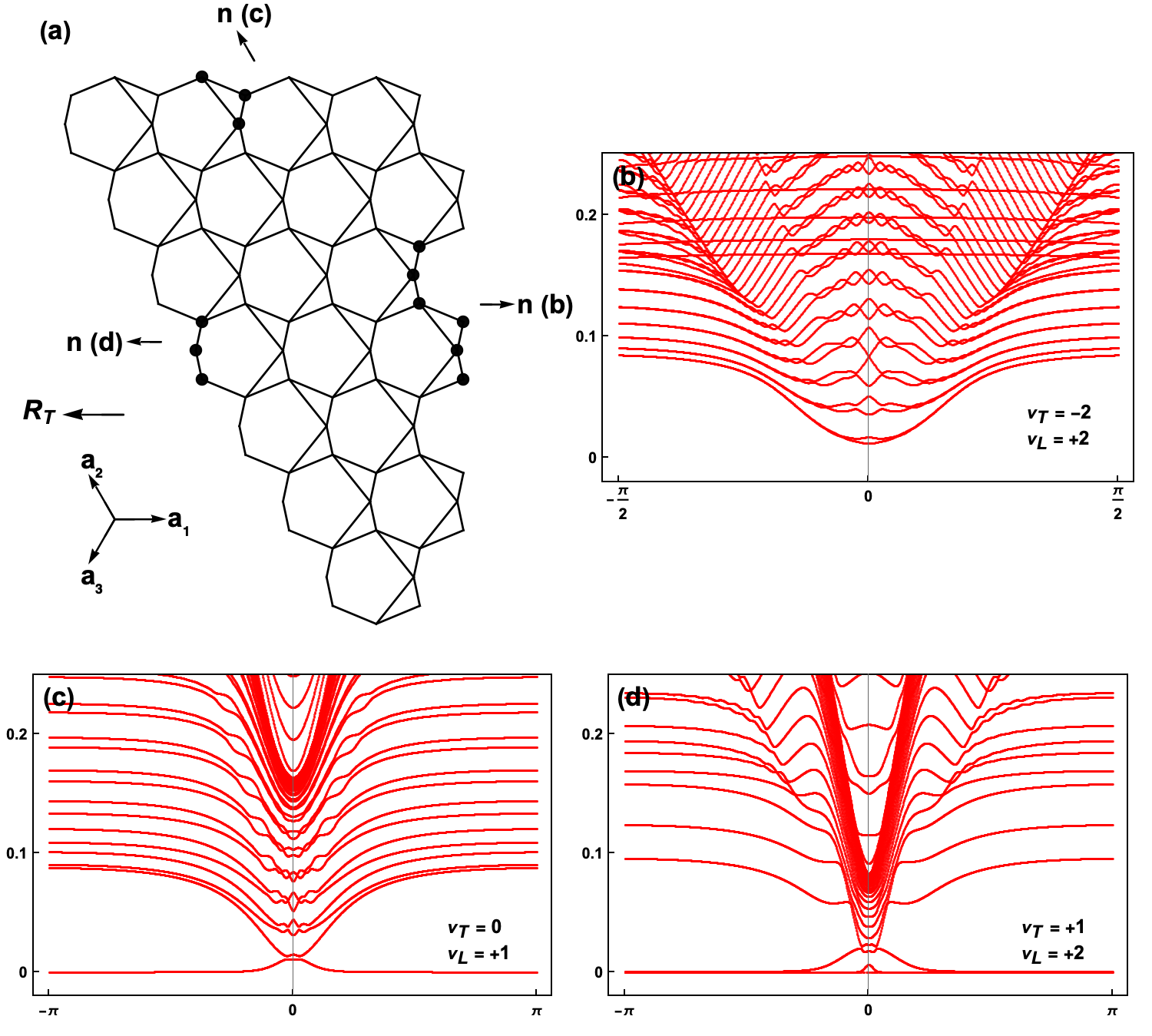}
    \caption{(a) The distorted kagom\'{e} lattice induced by SLC, 
                 under three different boundary terminations. 
                 For each boundary, the corresponding boundary unit cell is indicated by black circles.
             (b-d) The phonon spectrum of cylinders (or oblique cylinders) oriented along the $\mathbf{n}$-axis, 
             where one edge adopts the free boundary condition corresponding to the boundary 
             in Fig.~\ref{fig:boundary_modes}(a), while the opposite edge is clamped, 
             for ${\frac{J\gamma}{k}=-0.0125}$}.
    \label{fig:boundary_modes}
\end{figure}

\noindent{\it Topological floppy modes.}---For a finite sample cut 
from this periodic Maxwell lattice, 
imposing free boundary conditions necessarily removes a number of 
bonds proportional to the linear system size.
Since the distorted lattice under PBCs possesses only two bulk zero modes, 
the additional zero modes in the finite system must be localized at the boundaries. 
These boundary modes, while sensitive to the details of boundary terminations, 
remain robust against local perturbations. 
This topological character is analogous to 
that of time-reversal symmetric topological insulators, 
where symmetry-protected edge states arise 
due to the bulk-boundary correspondence. 
Based on this analogy, the boundary zero modes 
per unit cell can be categorized into topological type and local type, 
${\nu=\nu_T +\nu_L}$~\cite{KaneLubensky2014}. 

$\nu_L$ is a local count and is determined by the Maxwell Index theorem, 
while $\nu_T$ is a topological count that depends on the topological structure 
of the gapped phases. 
The local count $\nu_L$ depends on the boundary termination, 
which is calculated by the dipole moment per unit cell
$\nu_L = \frac{\mathbf{G}\cdot \mathbf{R}_L}{2\pi}$, 
    where $\mathbf{G}$ is the reciprocal lattice vector normal to the boundary, 
    and the local polarization $\mathbf{R}_L$ is given by
$\mathbf{R}_L = d\sum_{\text{sites }\mu}\mathbf{r}_\mu - \sum_{\text{bonds }\beta}\mathbf{r}_\beta$,
with $\mathbf{r}_\mu$ ($\mathbf{r}_\beta$) denoting the position of site $\mu$ 
(the center of bond $\beta$) within the unit cell.
The topological count $\nu_T$ can be expressed as  
\begin{equation}
\nu_{T}  = \frac{\mathbf{G} \cdot \mathbf{R}_T}{2\pi},
\label{eq:T}
\end{equation}
where topological polarization ${\mathbf{R}_T = \sum_i n_i \mathbf{a}_i}$ 
is a generalization of the winding number in one dimension. 
The integers $n_i$ are defined as  
\begin{equation}
n_i = \frac{1}{2\pi i} \oint_{C_i} d\mathbf{q} \cdot \mathrm{Tr}\!\left[ Q(\mathbf{q})^{-1} \nabla_{\mathbf{q}} Q(\mathbf{q}) \right],
\label{eq:ni_trace}
\end{equation}
with $Q(\mathbf{q})$ being the equilibrium matrix and $C_i$ being closed loops 
in the Brillouin zone 
along the reciprocal vectors $\mathbf{B}_i$. For the distorted lattice discussed above, 
the corresponding topological polarization is ${\mathbf{R}_{T} = -\mathbf{a}_1}$, 
which realizes a topological phase of the deformed kagom\'e lattice.

Fig.~\ref{fig:boundary_modes}(a) shows the distorted lattice induced by SLC, 
under three different boundary terminations. 
For each boundary, the corresponding boundary unit cell is indicated by black circles.
Figs.~\ref{fig:boundary_modes}(b)–(d) display the phonon spectra of cylinders (or oblique cylinders) 
oriented along the $\mathbf{n}$-axis, where one edge adopts the free boundary condition 
corresponding to the boundary in Fig.~\ref{fig:boundary_modes}(a), 
while the opposite boundary is clamped. 
This ensures that the observed zero modes arise solely from the free boundary. 
The number of zero modes per boundary unit cell agrees well with 
the counts from $\nu_T +\nu_L$. 
The lattice distortion used in Fig.~\ref{fig:boundary_modes} is determined under PBCs. 
With free boundaries, the spin configuration and the site displacements near the boundary 
generally deviate from their bulk ones. 
Equivalently, boundary introduce a local perturbation to the boundary geometry 
relative to the periodic bulk distortion. 
After perturbing the boundary sites, the phonon spectrum 
remains similar to Figs.~\ref{fig:boundary_modes}(b)--(d), 
and the boundary floppy modes remain unchanged.
This provides a direct demonstration of the robustness that 
is protected by Maxwell index theorem and bulk topology.

The results above establish that the SLC can induce a topological Maxwell lattice 
with nontrivial topological floppy modes. Different spin orders selected in other parameter regimes can, 
in turn, lead to other phonon phases. 
For the NN limit of Eq.~\eqref{eq:H_eff}, when ${1/6 < b < 1/3}$, 
the ground states exhibit a large macroscopic degeneracy, 
with magnetization per site ranging from $-1/9$ to $1/9$ \cite{PhysRevB.105.174424}. 
Applying an external magnetic field can lift this degeneracy and stabilize the $1/9$ plateau. 
In particular, the lattice distortion associated with the $1/9$ plateau can remove SSSs.
The corresponding phonon spectrum has the topological polarization $\mathbf{R}_{T} = 0$, 
belonging to a trivial phase with only local boundary floppy modes (see the Appendix). 
The fact that different spin states can realize distinct topological floppy modes 
suggests the possibility of manipulating and controlling the associated topological 
mechanical responses via the spin states.

\noindent{\it Discussion.}---Our study establishes SLC as a natural microscopic route
 for realizing topological mechanical phenomena in crystalline materials, and this 
 mechanism is dubbed ``magnetopological mechanics''. 
 Through the kagom\'e Heisenberg model, we demonstrated that strong SLC can 
 spontaneously distort the lattice, remove SSSs, and yield a topological Maxwell lattice 
 endowed with a nontrivial topological polarization. The resulting topological floppy modes, 
 previously achieved only in designed spring-mass systems, thus emerges intrinsically 
 from the interplay between spin and lattice degrees of freedom.

Beyond the specific example studied here, this framework
of ``magnetopological mechanics''
suggests a broader unifying principle linking magnetic order 
and topological floppy modes. 
Different spin configurations correspond to distinct lattice distortions 
and hence to different topological phases of the phonon spectrum. 
The domain wall of two symmetry-related spin states could host the topological 
floppy modes. 
Several extensions follow naturally from our results. 
One is the realization of mechanical Weyl models in magnetic materials, 
such as two-dimensional deformed square lattices~\cite{PhysRevLett.116.135503} 
or higher-dimensional generalized pyrochlore lattices~\cite{PhysRevLett.117.068001}, 
where lattice geometry and constraints create isolated zero-frequency points 
and topologically protected bulk and surface modes. 
The pyrochlore could immediately find applications in the Cr-based spinels 
where the SLC, spin states and lattice distortion were extensively 
studied~\cite{PhysRevLett.93.197203,PhysRevB.74.134409,PhysRevLett.88.067203,PhysRevB.66.064403}, 
but the topological aspects of phonons are not yet explored~\cite{inpreparation}. 
Another is the realization of self-dual mechanical lattices~\cite{Fruchart2020,PhysRevLett.127.018001}, 
including twisted kagom\'e structures, which feature a doubly degenerate vibrational
spectrum and non-Abelian geometric phases. In fact, twisted kagom\'e structures
are realized in many rare-earth magnets such as HoAgGe and CePdAl~\cite{Zhao_2020,Zhao_2019}.
Both classes could, in principle, be engineered via SLC-induced lattice distortions. 
More generally, beyond spin orders, other electronic orders 
may also drive crystalline 
lattices into distorted lattices, potentially hosting topological phonon modes, 
Weyl points or lines, and self-dual mechanical lattices.

Finally, the domain wall gapless phonons in Maxwell lattice quantum materials 
and the bulk gapless topological phonons, whether induced by the lattice structure itself 
or through SLC or interactions with other degrees of freedom, 
are quite soft due to their topological origin. 
The soft domain-wall or bulk phonon could 
enhance electron-phonon coupling and may give rise to high-temperature superconductivity. 
In fact, interfacial phonons were proposed to be one possible ingredient for the very high
superconductivity transition temperature on the FeSe/SrTiO$_3$ interface~\cite{Ge_2014}. 
The mechanism of superconductivity driven by such topological phonons warrants further investigation.
This possibility can even be
relevant to square lattice and cubic lattice quantum materials. Although these are not typical geometrically frustrated systems, they are Maxwell lattices. Many
actively explored platforms, including cuprates, Fe-based
superconductors, and perovskite systems, belong to this
broader class. Their possible topological phonons, and
especially their coupling to electronic degrees of freedom, warrant systematic investigation.
Overall, our work bridges quantum materials and topological floppy modes, 
demonstrating that topological phonon floppy modes can naturally emerge 
in realistic quantum materials and opening pathways for designing, realizing,
and exploring the emerging phonon-related physics of Maxwell lattice quantum materials.

\noindent\emph{Acknowledgments.}---This work is supported by NSFC with Grants No.~92565110 
and No.~12574061, by the Ministry of Science and Technology of China with Grants No.~2021YFA1400300.

\bibliography{ref}

\clearpage            
\onecolumngrid        

\appendix
\section{Spin configurations of Ising-type effective Hamiltonian}
\begin{figure}[htbp]
    \centering
    \subfloat{%
        \begin{tikzpicture}
            \node[anchor=north west, inner sep=0] (img) at (0,0)
                {\includegraphics[width=0.3\columnwidth]{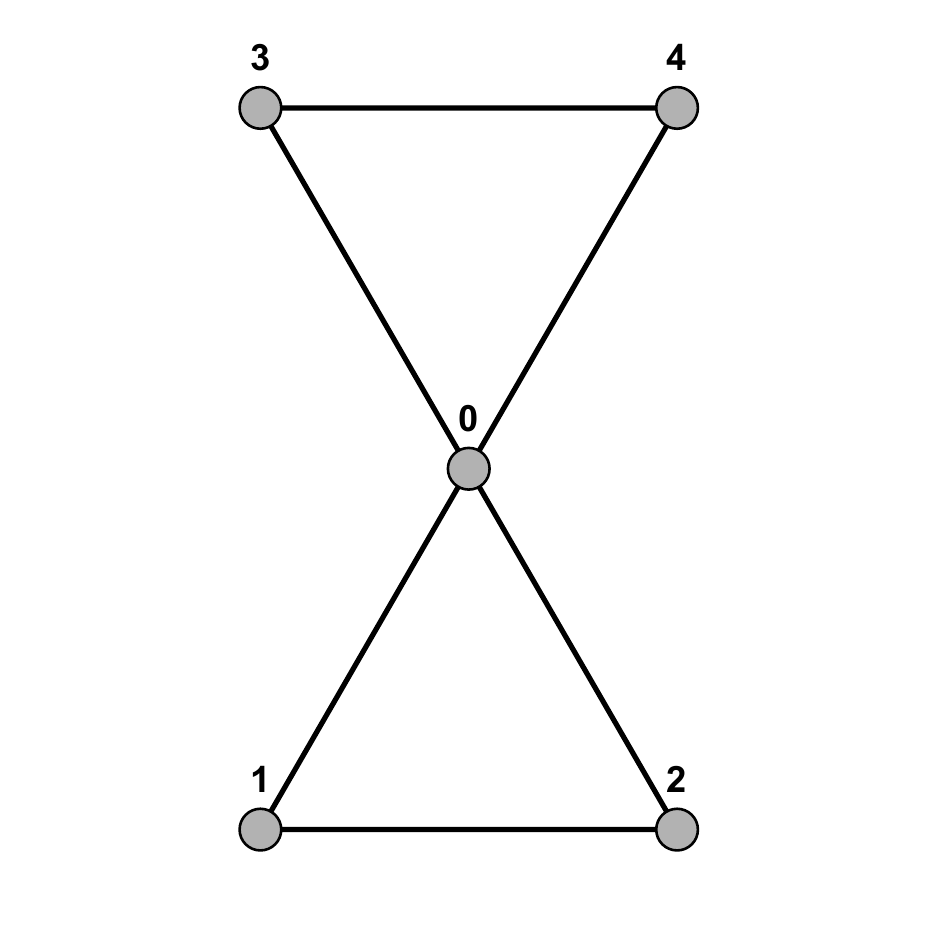}};
            \node[anchor=north west] at ([xshift=1pt,yshift=8pt]img.north west)
                {\fontsize{6}{12}\selectfont\textbf{(a)}};
        \end{tikzpicture}%
        \label{fig:SM_pattern}%
    }%
    \hspace{0.5em}
    \subfloat{%
        \begin{tikzpicture}
            \node[anchor=north west, inner sep=0] (img) at (0,0)
                {\includegraphics[width=0.48\columnwidth]{local_spin_configuration.pdf}};
            \node[anchor=north west] at ([xshift=-8pt,yshift=0pt]img.north west)
                {\fontsize{6}{12}\selectfont\textbf{(b)}};
        \end{tikzpicture}%
        \label{fig:SM_local_pattern}%
    }%
    \caption{   
        (a) Local spin pattern for site $i$.
        (b)Representative local spin configurations on the kagome lattice
        used to evaluate the energy of $\mathcal{H}_{e}^{\mathrm{Ising}}$.
        Red circles indicate spin up ($+\hat{z}$),
        while blue circles indicate spin down ($-\hat{z}$).
        All these configurations, together with their symmetry-related counterparts,
        act as ``Lego blocks'',
        from which arbitrary global spin configurations can be constructed.}
    \label{fig:SM_1}
\end{figure}

In the nearest-neighbor (NN) limit, the Ising-type effective Hamiltonian is
\begin{equation}
    \mathcal{H}_{N}^{\mathrm{Ising}}
    =  \sum_{\langle i,j \rangle} \left(J - \frac{Jb}{2}\right) \sigma_i \sigma_j
    + \sum_{\langle i,j \rangle_{2}}  \frac{Jb}{2} \sigma_i \sigma_j + \sum_{\langle i,j \rangle_{3\parallel}} Jb \,  \sigma_i \sigma_j
    \label{eq:SM_H_nearest_Ising}
\end{equation}
the ground state configurations of this Hamiltonian can be determined by several local spin patterns, as illustrated in Fig.~\ref{fig:SM_local_pattern}. All these patterns, together with their symmetry-related counterparts, act as ``Lego blocks'',
from which arbitrary global configurations on the kagome lattice can be constructed. To relate the energy of a global configuration to the energies of these local patterns, it is convenient to rewrite the Hamiltonian as a sum over local patterns. 
We emphasize that such a decomposition is not unique: the same Hamiltonian can be partitioned into local terms in different ways, depending on how one distributes the bond energies among patterns.  
For example, the Hamiltonian above can be decomposed as
\begin{subequations}\label{eq:SM_H_nearest_Ising_all}
\begin{equation}\label{eq:SM_H_nearest_Ising1}
\begin{aligned}
\mathcal{H}_{1}^{\mathrm{Ising}}
= \sum_{ i } & \frac{1}{3}\Bigl(J-\frac{Jb}{2}\Bigr)\Big(
\sigma_{i,0}\sigma_{i,1} + \sigma_{i,1}\sigma_{i,2} + \sigma_{i,2}\sigma_{i,0}
+ \sigma_{i,0}\sigma_{i,3} + \sigma_{i,3}\sigma_{i,4} + \sigma_{i,4}\sigma_{i,0}
\Big) \\
&+ \frac{Jb}{2} \Big( \sigma_{i,1}\sigma_{i,3} + \sigma_{i,2}\sigma_{i,4} \Big)
+ Jb \Big( \sigma_{i,1}\sigma_{i,4} + \sigma_{i,2}\sigma_{i,3} \Big) \, ,
\end{aligned}
\end{equation}

\begin{equation}\label{eq:SM_H_nearest_Ising2}
\mathcal{H}_{2}^{\mathrm{Ising}}
= \sum_{ i } \Bigl(J-\frac{Jb}{2}\Bigr)\Big( \sigma_{i,1}\sigma_{i,2} + \sigma_{i,3}\sigma_{i,4} \Big)
+ \frac{Jb}{2} \Big( \sigma_{i,1}\sigma_{i,3} + \sigma_{i,2}\sigma_{i,4} \Big)
+ Jb \Big( \sigma_{i,1}\sigma_{i,4} + \sigma_{i,2}\sigma_{i,3} \Big) \, ,
\end{equation}

\begin{equation}\label{eq:SM_H_nearest_Ising3}
\begin{aligned}
\mathcal{H}_{3}^{\mathrm{Ising}}
= \sum_{ i } & \frac{J}{3} \Big(
\sigma_{i,0}\sigma_{i,1} + \sigma_{i,1}\sigma_{i,2} + \sigma_{i,2}\sigma_{i,0}
+ \sigma_{i,0}\sigma_{i,3} + \sigma_{i,3}\sigma_{i,4} + \sigma_{i,4}\sigma_{i,0}
\Big) \\
&-\frac{Jb}{2} \Big( \sigma_{i,1}\sigma_{i,2} + \sigma_{i,3}\sigma_{i,4} \Big)
+ \frac{Jb}{2} \Big( \sigma_{i,1}\sigma_{i,3} + \sigma_{i,2}\sigma_{i,4} \Big)
+ Jb \Big( \sigma_{i,1}\sigma_{i,4} + \sigma_{i,2}\sigma_{i,3} \Big) \, .
\end{aligned}
\end{equation}
\end{subequations}
$\sigma_{i,0}$--$\sigma_{i,4}$ are spin operators belonging to pattern $i$, as shown in Fig.~\ref{fig:SM_pattern}.
The corresponding energies of the local spin patterns for each decomposition are 
\begin{equation}
\begin{aligned}
\begin{aligned}
& E^{(\mathrm{I})}_{1}   = \frac{2}{3}J-\frac{1}{3}Jb,\\
& E^{(\mathrm{II})}_{1}  = -\frac{2}{3}J+\frac{10}{3}Jb,\\
& E^{(\mathrm{III})}_{1} = -\frac{2}{3}J+\frac{1}{3}Jb,\\
& E^{(\mathrm{IV})}_{1}  = -\frac{2}{3}J-\frac{2}{3}Jb,\\
& E^{(\mathrm{V})}_{1}   = -\frac{2}{3}J+\frac{4}{3}Jb,\\
& E^{(\mathrm{VI})}_{1}  = \frac{2}{3}J-\frac{10}{3}Jb.
\end{aligned}
\qquad
\begin{aligned}
& E^{(\mathrm{I})}_{2}   = 0,\\
& E^{(\mathrm{II})}_{2}  = 2J+2Jb,\\
& E^{(\mathrm{III})}_{2} = 0,\\
& E^{(\mathrm{IV})}_{2}  = -2J,\\
& E^{(\mathrm{V})}_{2}   = -2J+2Jb,\\
& E^{(\mathrm{VI})}_{2}  = 2J-4Jb.
\end{aligned}
\qquad
\begin{aligned}
& E^{(\mathrm{I})}_{3}   = \frac{2}{3}J,\\
& E^{(\mathrm{II})}_{3}  = -\frac{2}{3}J+2Jb,\\
& E^{(\mathrm{III})}_{3} = -\frac{2}{3}J,\\
& E^{(\mathrm{IV})}_{3}  = -\frac{2}{3}J,\\
& E^{(\mathrm{V})}_{3}   = -\frac{2}{3}J+2Jb,\\
& E^{(\mathrm{VI})}_{3}  = \frac{2}{3}J-4Jb.
\end{aligned}
\end{aligned}
\end{equation}
Since all three decompositions reproduce the same total Hamiltonian, they must give the same total energy for any spin configuration. 
Suppose that a candidate eigenstate (or ground state manifold) can be realized by a tiling of the lattice using a set of local patterns $\{i\}$, with each pattern $i$ appearing with a frequency $x_i$. 
Then the total energy computed from any decomposition must coincide, which imposes the consistency condition
\begin{equation}
\sum_i x_i\,E^{(i)}_{1}
= \sum_i x_i\,E^{(i)}_{2}
= \sum_i x_i\,E^{(i)}_{3},
\label{eq:energy_consistency}
\end{equation}
where $E^{(i)}_{\alpha}$ denotes the energy of pattern $i$ under the $\alpha$th decomposition.
In particular, a single pattern cannot tile the lattice: if one pattern $i$ alone were sufficient, Eq.~\eqref{eq:energy_consistency} would require
$E^{(i)}_{1}=E^{(i)}_{2}=E^{(i)}_{3}$, which is not satisfied for any of the patterns listed above.

Among these decompositions, $\mathcal{H}^{\mathrm{Ising}}_{3}$ is the most convenient for identifying the ground state configuration, because it has only two energetically lowest-energy local patterns type III and type IV for $1/6<b<1/3$ and can tile the kagome lattice. 
Moreover, Eq.~\eqref{eq:energy_consistency} fixes their ratio 
$x_{\mathrm{III}}:x_{\mathrm{IV}} = 2:1$.

When $J_{3*}=0$ and $J_{4}=0$, the Ising-type effective Hamiltonian is
\begin{equation}
    \mathcal{H}_e^{\mathrm{Ising}}
    =  \sum_{\langle i,j \rangle} \left(J - \frac{Jb}{2}\right) \sigma_i \sigma_j
    + \sum_{\langle i,j \rangle_{2}} \left(J_2 + \frac{Jb}{2}\right) \sigma_i \sigma_j
    + \sum_{\langle i,j \rangle_{3\parallel}} \left(J_{3\parallel} + Jb \right) \sigma_i \sigma_j \, .
    \label{eq:SM_H_extended_Ising}
\end{equation}
We still adopt the third decomposition introduced above. The corresponding energies of patterns are
\begin{equation}
\begin{aligned}
E^{(\mathrm{I})}   &= \frac{2}{3}J,\\
E^{(\mathrm{II})}  &= 2J_2+2J_{3\parallel}+2Jb-\frac{2}{3}J,\\
E^{(\mathrm{III})} &= -\frac{2}{3}J,\\
E^{(\mathrm{IV})}  &= 2J_2-2J_{3\parallel}-\frac{2}{3}J,\\
E^{(\mathrm{V})}   &= -2J_2+2J_{3\parallel}+2Jb-\frac{2}{3}J,\\
E^{(\mathrm{VI})}  &= -2J_2-2J_{3\parallel}-4Jb+\frac{2}{3}J.
\end{aligned}
\label{eq:pattern_energies_extended}
\end{equation}
For $1/6<b<1/3$, the two lowest-energy local patterns remain type III and type IV. 
Which one is lower depends on the small further-neighbor interactions $J_2$ and $J_{3\parallel}$. 
However, since $J_2$ and $J_{3\parallel}$ are much smaller than $J$ and $b$, the ground state configuration is still built from patterns type III and type IV with ratio $x_{\mathrm{III}}:x_{\mathrm{IV}} = 2:1$.

\section{Distorted lattice and topological polarization of the 1/9 Plateau phase}
\label{appendix}

In the nearest-neighbor (NN) limit, the effective spin Hamiltonian is
\begin{equation}
    \mathcal{H}_{\mathrm{eff}}
    = J \sum_{\langle i,j \rangle_1} 
      \big[ \mathbf{S}_i \cdot \mathbf{S}_j 
      - b\, (\mathbf{S}_i \cdot \mathbf{S}_j)^2 \big]
      + \mathcal{H}_{\mathrm{FN}},
    \label{eq:SM_H_eff}
\end{equation}
where $b = \frac{J \gamma^2}{k}$ is the dimensionless spin lattice coupling (SLC) strength.
The three-spin quartic term $\mathcal{H}_{\mathrm{FN}}$ takes the form
\begin{equation}
    \mathcal{H}_{\mathrm{FN}}
    = - \frac{Jb}{2} 
      \sum_i \sum_{j \neq k \in \mathcal{N}(i)} 
      (\hat{\mathbf{e}}_{ij} \cdot \hat{\mathbf{e}}_{ik})
      (\mathbf{S}_i \cdot \mathbf{S}_j)
      (\mathbf{S}_i \cdot \mathbf{S}_k).
    \label{eq:SM_H_FN}
\end{equation}

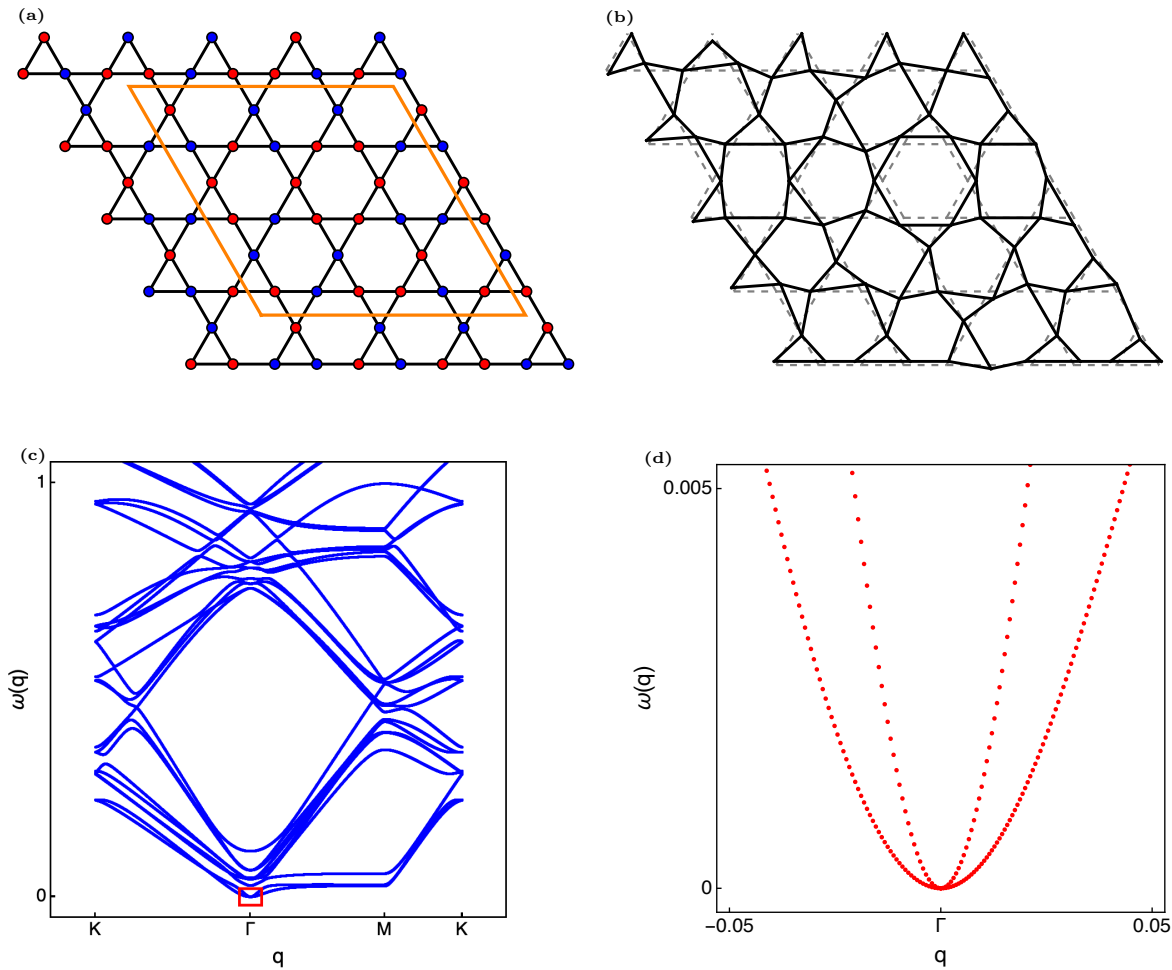
\begin{figure}[htbp]
    \centering

    \subfloat{%
        \begin{tikzpicture}
            \node[anchor=north west, inner sep=0] (img) at (0,0)
                {\includegraphics[width=0.43\columnwidth]{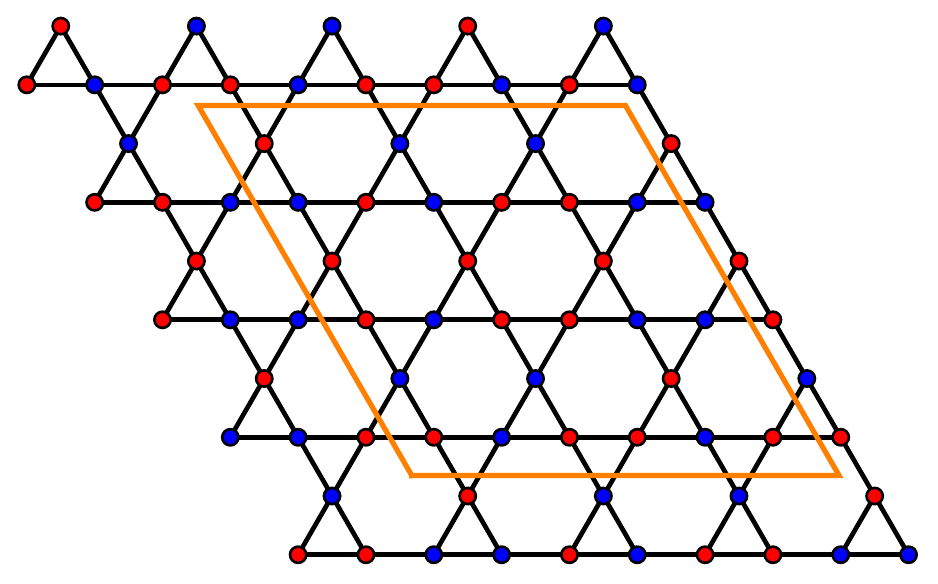}};
            \node[anchor=north west] at ([xshift=1pt,yshift=8pt]img.north west)
                {\fontsize{6}{12}\selectfont\textbf{(a)}};
        \end{tikzpicture}%
        \label{fig:SM_conf}%
    }%
    \vspace{0.6em}
    \subfloat{%
        \begin{tikzpicture}
            \node[anchor=north west, inner sep=0] (img) at (0,0)
                {\includegraphics[width=0.43\columnwidth]{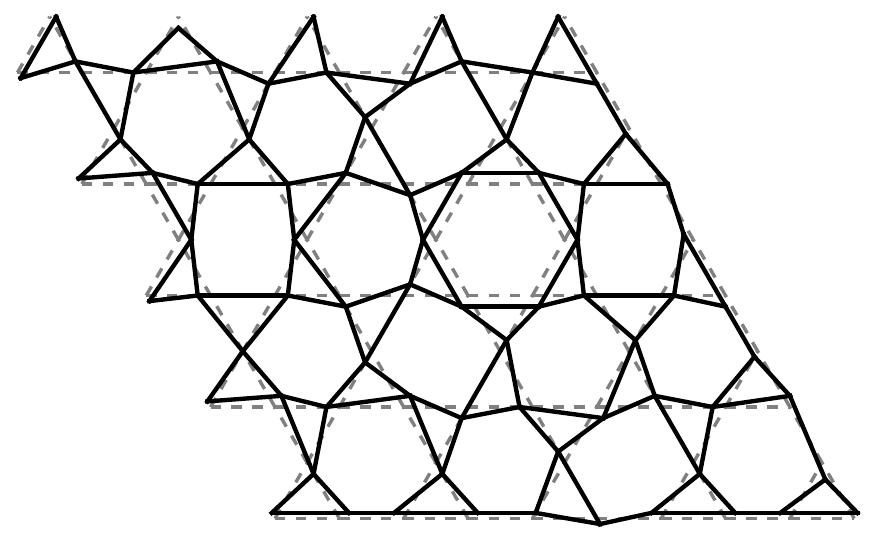}};
            \node[anchor=north west] at ([xshift=1pt,yshift=8pt]img.north west)
                {\fontsize{6}{12}\selectfont\textbf{(b)}};
        \end{tikzpicture}%
        \label{fig:SM_lattice}%
    }%

    \vspace{0.6em}

    \subfloat{%
        \begin{tikzpicture}
            \node[anchor=north west, inner sep=0] (img) at (0,0)
                {\includegraphics[width=0.37\columnwidth]{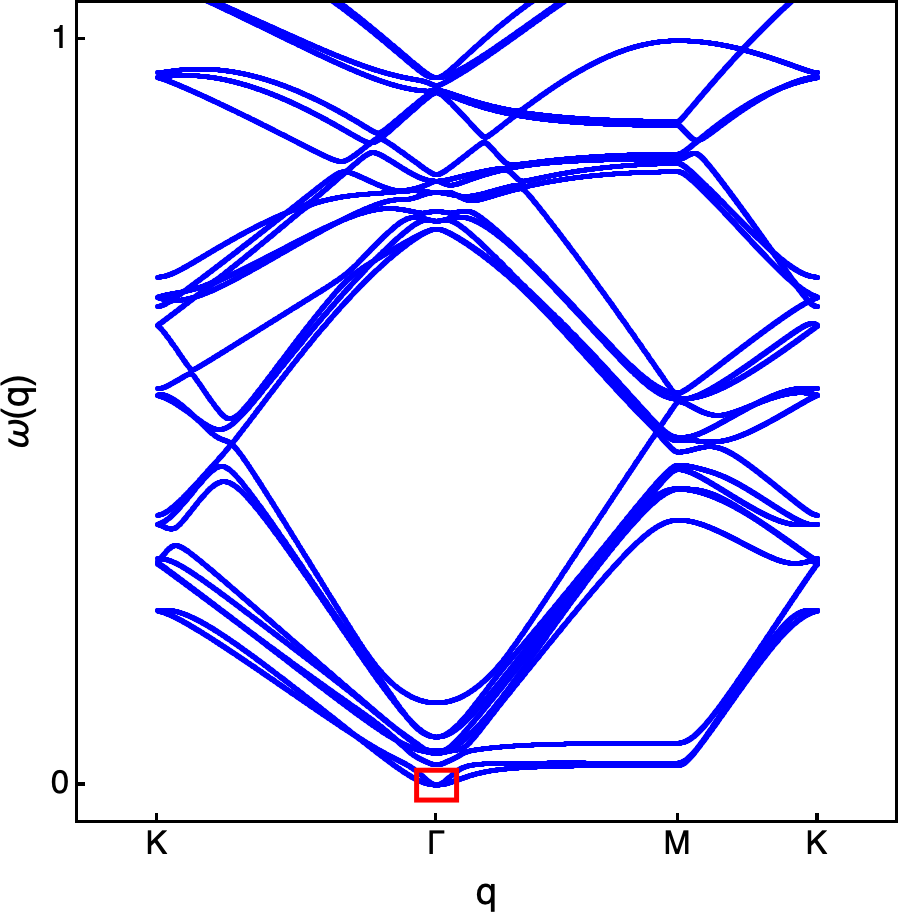}};
            \node[anchor=north west] at ([xshift=1pt,yshift=8pt]img.north west)
                {\fontsize{6}{12}\selectfont\textbf{(c)}};
        \end{tikzpicture}%
        \label{fig:SM_phonon}%
    }%
    \hspace{4.8em}
    \subfloat{%
        \begin{tikzpicture}
            \node[anchor=north west, inner sep=0] (img) at (0,0)
                {\includegraphics[width=0.4\columnwidth]{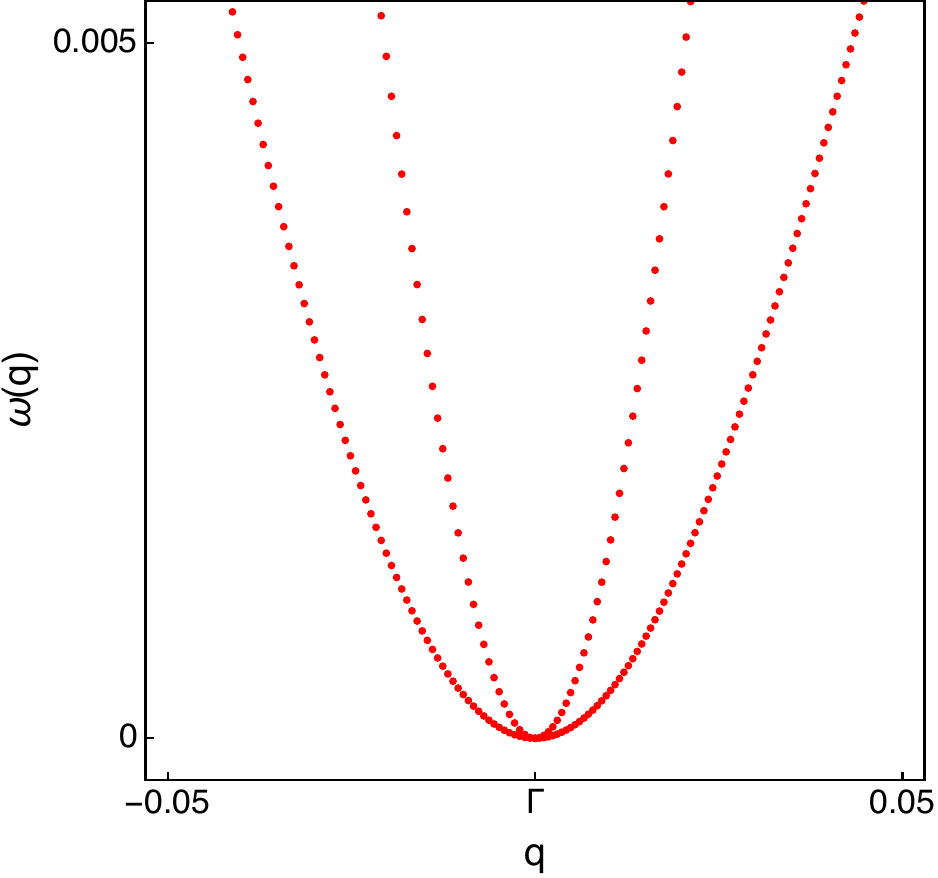}};
            \node[anchor=north west] at ([xshift=1pt,yshift=8pt]img.north west)
                {\fontsize{6}{12}\selectfont\textbf{(d)}};
        \end{tikzpicture}%
        \label{fig:SM_phononmin}%
    }%

    \caption{
        (a) $1/9$-plateau spin configuration of the Ising Hamiltonian~\eqref{eq:SM_H_Ising}. Red (blue) circles denote spins pointing along $+\hat{z}$ ($-\hat{z}$). The orange parallelogram marks the unit cell.
        (b) Corresponding lattice distortion induced by the SLC.
        (c) Phonon spectrum of the distorted lattice along $K\!-\!\Gamma\!-\!M\!-\!K$.
        (d) phonon spectrum of the red box region in (c).
}
    \label{fig:SM_ex1}
\end{figure}

In the weak SLC region ($0 < b < \frac{1}{6}$), the ground state remains the coplanar $120^{\circ}$ spin configuration, identical to that of the NN Heisenberg model~\cite{PhysRevB.105.174424}.
All NN spin bond energy are equal in this phase, hence the optimal displacements vanish, $\mathbf{u}_i^{*}=0$, and the lattice stays undistorted.

In contrast, in the strong-SLC regime ($b>1/6$), the second term in Eq.~\eqref{eq:SM_H_eff} favors collinear spins. 
Thus the effective Hamiltonian can be mapped onto an Ising-type model by setting $\mathbf{S}_i = \hat{z}\,\sigma_i$ with $\sigma_i = \pm 1$:
\begin{equation}
    \mathcal{H}^{\mathrm{Ising}}
    = \left(J - \frac{Jb}{2}\right)
        \sum_{\langle i,j \rangle_1} \sigma_i \sigma_j
    + \frac{Jb}{2} 
        \sum_{\langle i,j \rangle_2} \sigma_i \sigma_j
    + Jb \,
        \sum_{\langle k,l \rangle_{3\parallel}} \sigma_k \sigma_l,
    \label{eq:SM_H_Ising}
\end{equation}
where $\langle i,j\rangle_1$, $\langle i,j\rangle_2$, and $\langle k,l\rangle_{3\parallel}$ represent the NN, next-nearest-neighbor, and bond-directional third-nearest-neighbor interactions respectively.

For $1/6<b<1/3$ the lowest-energy patterns are types~$\mathrm{III}$ and~$\mathrm{IV}$ (and their symmetry-related counterparts). 
Arbitrary assemblies of these patterns generate a macroscopically degenerate ground state with magnetization per site in $[-1/9,\,1/9]$. 
An small external magnetic field lifts the degeneracy and stabilizes the $1/9$ plateau (Fig.~\ref{fig:SM_conf}), whose SLC-induced distortion is shown in Fig.~\ref{fig:SM_lattice}. 
The resulting distorted lattice has a 27-site, 54-bond unit cell (outlined by the orange parallelogram in Fig.~\ref{fig:SM_conf}).

Using this distorted lattice, we compute the bulk phonon spectrum under periodic boundary conditions following Ref.~\cite{doi:10.1073/pnas.1119941109}. 
As shown in Figs.~\ref{fig:SM_phonon} and~\ref{fig:SM_phononmin}, the distortion  gaps out the bulk zero modes along the $\Gamma$--$M$ direction, leaving only the two trivial translational zero modes at $q=0$. 
The lattice is therefore isostatic. 
Moreover, the topological polarization is $\mathbf{R}_T=0$, implying that the $1/9$-plateau lattice supports only local boundary floppy modes.

\end{document}